\documentclass[prl,twocolumn,showpacs]{revtex4}
\usepackage{epsfig}
\usepackage{dcolumn}
\usepackage{color}

\begin{document}
\bibliographystyle{apsrev}

\title{Scanning Tunneling Spectroscopy Study of the Proximity Effect 
in a Disordered Two-Dimensional Metal}

\author{L. Serrier-Garcia$^1$}
\author{J. C. Cuevas$^2$}
\author{T. Cren$^1$}
\email{tristan.cren@upmc.fr}
\author{C. Brun$^1$}
\author{V. Cherkez$^1$}
\author{F. Debontridder$^1$}
\author{D. Fokin$^{1,3}$}
\author{F. S. Bergeret$^{4}$}
\author{D. Roditchev$^1$}

\affiliation{$^1$Institut des Nanosciences de Paris, Universit\'{e} Pierre
et Marie Curie (UPMC) and CNRS-UMR 7588, 4 place Jussieu, 75252 Paris, France}

\affiliation{$^2$Departamento de F\'{\i}sica Te\'orica de la Materia Condensada, 
Universidad Aut\'onoma de Madrid, E-28049 Madrid, Spain}

\affiliation{$^3$Joint Institute for High Temperatures RAS 125412, Moscow, Russia}

\affiliation{$^4$Centro de F\'{\i}sica de Materiales (CFM-MPC), Centro Mixto CSIC-UPV/EHU and
Donostia International Physics Center (DIPC), P$^{\rm o}$ Manuel de Lardizabal 4, 
E-20018 San Sebasti\'an, Spain}
\date{\today}

\begin{abstract}
The proximity effect between a superconductor and a highly diffusive two-dimensional metal 
was revealed in a Scanning Tunneling Spectroscopy experiment. The \textit{in-situ} elaborated 
samples consisted of superconducting single crystalline Pb islands interconnected by a 
non-superconducting atomically thin disordered Pb wetting layer. In the vicinity of each superconducting 
island the wetting layer acquires specific tunneling characteristics which reflect the 
interplay between the proximity-induced superconductivity and the inherent electron 
correlations of this ultimate diffusive two-dimensional metal. The observed spatial 
evolution of the tunneling spectra was accounted for theoretically by combining the 
Usadel equations with the theory of dynamical Coulomb blockade; the relevant length and 
energy scales were extracted and found in agreement with available experimental data.
\end{abstract}


\maketitle

When a normal metal (N) is in close proximity to a superconductor (S), it acquires
genuine superconducting properties which are reflected in the modification of its local 
density of states (DOS). This phenomenon, known as \emph{proximity effect}, has been studied 
for decades \cite{deGennes1964} and is used nowadays in various superconducting quantum 
devices \cite{Cleuziou2006,Giazotto2010}. Most of the experimental studies of the proximity 
effect have focused on the analysis of three-dimensional diffusive metals \cite{Pannetier2000}, 
whereas reports on two-dimensional (2D) systems are scarce. Due to reduced screening, 
enhanced electron correlations, and localization at low temperatures, the proximity 
effect in 2D metals is expected to be drastically modified. However, not much is known about 
the interplay between superconducting correlations induced by proximity and electron correlations in disordered 
2D metals. The central goal of this work is to shed new light on this fundamental issue.

So far, apart from notable exceptions \cite{Gueron1996,Moussy2001,Escoffier2004,leSueur2008}, 
most of the reported experiments did not provide spatially-resolved information on the proximity 
effect. This is mainly due to the difficulties on applying Scanning Tunneling Microscopy/Spectroscopy 
(STM/STS) techniques to \emph{ex-situ} elaborated (and thus, surface-contaminated) mesoscopic 
systems. On the other hand, the progress in the controlled growth of atomically clean superconducting 
materials under ultrahigh vacuum has opened the possibility to probe the proximity effect in \emph{in-situ} 
STM/STS experiments with high spatial and energy resolution.  Very recently, Kim 
\emph{et al.} \cite{Kim2012} reported a STM/STS study in a system consisting of 5 monolayer (ML) 
thick superconducting Pb islands grown on Si(111) and an atomic overlayer of the striped 
incommensurate (SIC) phase of Pb. Unfortunately, the analysis of the proximity effect in the 
SIC layer was performed at relatively high temperatures ($T=4.3$ K) and thus, the energy 
resolution was rather limited. More importantly, this experiment did not reveal any specific 
feature related to the low-dimensionality of the SIC layer. Besides, SIC phase Pb/Si(111) becomes 
superconducting below $T_C = 1.8$ K and thus, it is not an ideal system where to study the 
proximity effect.

\begin{figure*}[t]
\includegraphics*[width=\textwidth,clip]{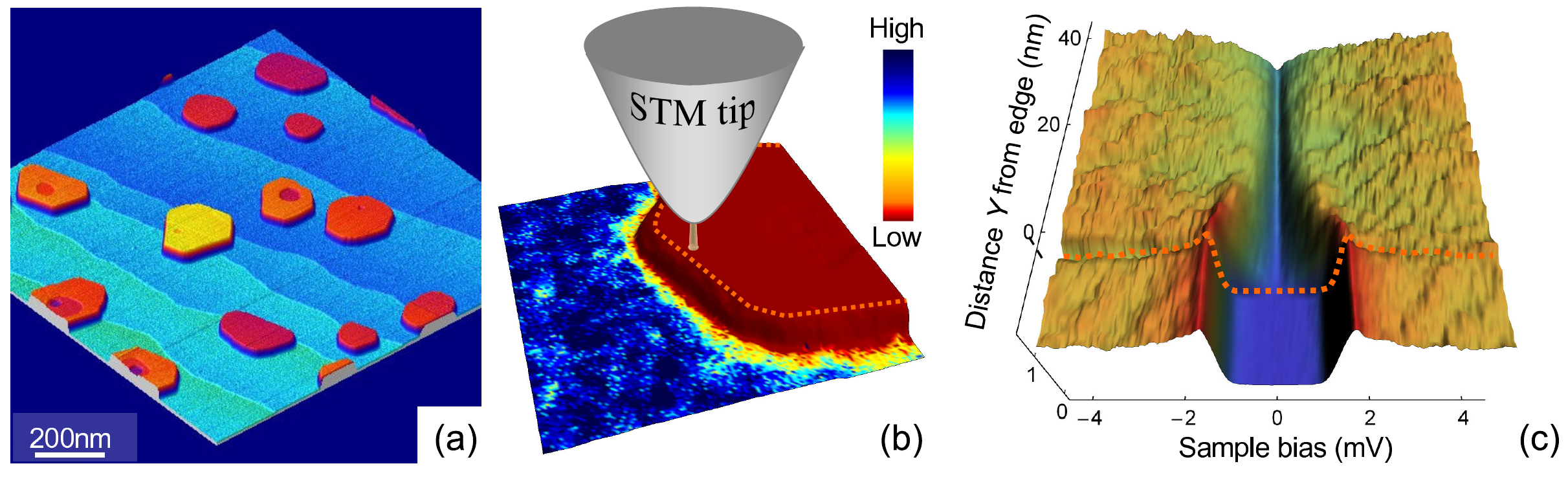}
\caption{(a) - topographic constant current ($I=0.1$ nA, $V=25$ mV) STM image : Pb islands and the WL (appearing in blue), cover single atomic terraces of underlying Si substrate. (b) - schematic 
representation of the STM/STS experiment with a real example of a color-coded STS tunneling conductance 
map $dI(V=0,\textit{\textbf{r}})/dV$ superimposed onto 3D topographic STM image. The colors correspond to the conductance variations from 0 nS ('Low') to 69 nS ('High'). The fully gaped spectra on the Pb island appear in red, the normal WL - in blue, and the proximity region - in yellow. The 
dashed orange line indicates the position of the flat top edge of the Pb island. (c) Spatial variations of 
the local $dI(V)/dV$ spectra (rainbow pallet 0-100 nS) as a function of the distance from the island edge. 
The dashed line corresponds to the spectrum taken right at the island edge.} \label{Fig1}
\end{figure*}

In this Letter we explore experimentally and theoretically the proximity effect in a highly diffusive atomically-thin metal.
The samples were grown \emph{in situ} by deposition at room temperature of a few atomic layers 
of Pb onto an atomically clean $7\times 7$ reconstructed surface of undoped Si(111). 
The topographic STM images in Fig.~\ref{Fig1}(a) revealed the resulting Pb network consisting 
of atomically flat 7-13 ML thick single nano-crystals of Pb, interconnected by a 1-2 ML thick
wetting layer (WL) of Pb \cite{Budde2000}. In Fig.~\ref{Fig1}(b) the experimental geometry is 
sketched; the regions of interest include the island edge and the neighboring part of the WL. 
The experiments were performed in ultra-high vacuum $P\approx5\times10^{-11}$~mbar and mechanically 
sharpened Pt/Ir tips were used. In order to resolve fine spectroscopic features, the STM/STS 
experiments were conducted at $320$ mK, \emph{i.e.} at $T\approx$ $T_C/20$ \cite{Brun2009}. 
Local tunneling conductance spectra $dI(V,\textit{\textbf{r}})/dV$ were derived from 
the raw $I(V,\textit{\textbf{r}})$ data.

Our main finding is illustrated in Fig.~\ref{Fig1}(c) where we show the evolution of local 
$dI(V,Y)/dV$ spectra as a function of the distance $Y$ from the island edge. The superconducting 
state in the islands is characterized by a fully opened gap $dI(V=0)/dV=0$ (bottom curve in 
Fig.~\ref{Fig1}(c)), which is spatially homogeneous over the island area (red-colored in 
the STS map Fig.~\ref{Fig1}(b)). On the WL side far from the islands, the zero-bias conductance 
is high, and the tunneling spectra in these regions exhibit a V-shaped dip at zero-bias (upper 
curve in Fig.~\ref{Fig1}(c)), which from now on will be referred to as a \emph{zero-bias anomaly} 
(ZBA). The absence of a clear gap is an evidence of the intrinsic non-superconducting character of 
the WL \cite{Non-supercond WL}. As the STM tip approaches the island edge, the tunneling spectra 
evolve gradually from the non-superconducting ZBA to the superconducting gap shape thus evidencing 
the proximity effect on a scale of several nanometers (detailed conductance maps 
are presented in supplementary material \cite{SM}).

\begin{figure}[t]
\includegraphics[width=\columnwidth,clip]{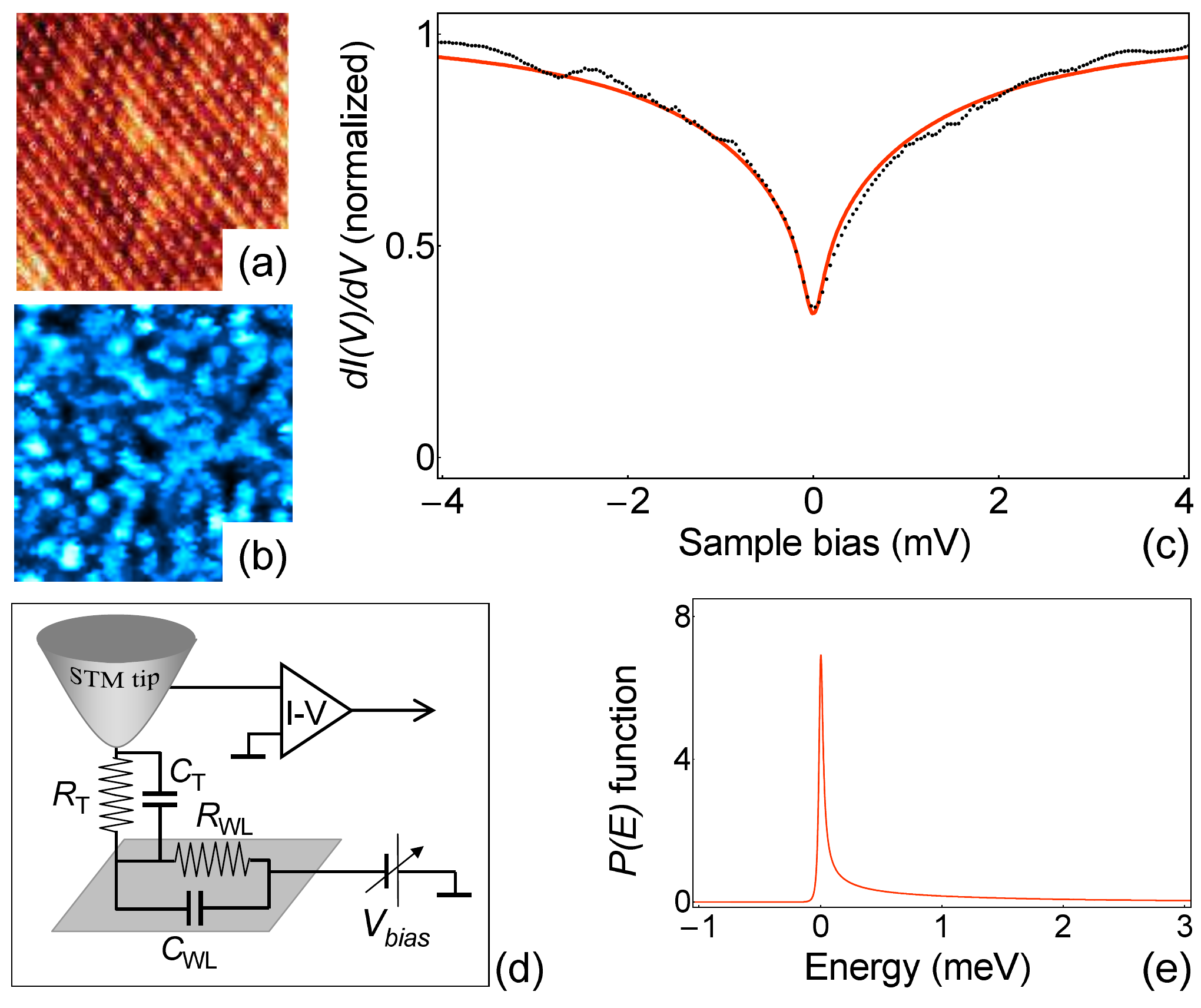}
\caption{(a) Topographic 5 nm$\times$5 nm STM image (taken at $I=0.1$ nA, $V=20$ mV) showing the atomic pattern 
at the top of a Pb island. (b) Topographic 30 nm$\times$30 nm STM image of the wetting layer.      
(c) dots - characteristic conductance spectrum measured in the WL away from the Pb island (normalized to unity at STS set point $I=0.4$ nA, $V=5$ mV). The 
solid line corresponds to the fit obtained using the DCB theory described in the text. (d) Considered 
equivalent circuit of the experiment (see text for details). (e) The $P(E)$ function used to 
obtain the fit shown in panel (c).} \label{Fig2}
\end{figure}

Let us first discuss the origin of the ZBA in the tunneling spectra of the WL which is, taking 
into account its thickness and the absence of conduction electrons at low temperatures in 
the undoped Si substrate, an ideal case of a disordered 2D conductor. Indeed, among the
reported Pb-Si systems, the atomically disordered Pb WL \cite{Budde2000} is characterized 
by an increased roughness in STM images ($\approx0.06$ nm in Fig.~\ref{Fig2}(b)) and lack 
of atomic ordering. This is at variance with the Pb islands which have smooth and perfectly 
ordered surfaces, see Fig.~\ref{Fig2}(a). In Fig.~\ref{Fig2}(c) we present a characteristic
$dI/dV$ tunneling spectrum taken in the WL far enough from the islands, to avoid the 
proximity effect. The shape of this ZBA resembles the diffusive anomalies found in tunneling 
junctions between disordered conductors \cite{Altshuler1985}. In those junctions, the suppressed
tunneling conductance at low bias is attributed to the reduction of the electron tunneling DOS
around the Fermi energy caused by the electron-electron interaction \cite{Altshuler1985}. 
Similar anomalies appear in ultrasmall tunnel junctions due to dynamical Coulomb blockade 
(DCB) \cite{Devoret1990,Girvin1990,Grabert1992,Brun DCB}, \emph{i.e.} due to the interaction of the tunneling electrons with the electromagnetic 
environment in which the junction is embedded \cite{Grabert1992}. Indeed, these two phenomena 
have been shown to be two sides of the same coin \cite{Nazarov1989,Ingold1992,Rollbuehler2001}. 
In particular, Rollb\"uhler and Grabert \cite{Rollbuehler2001} have demonstrated that the tunneling 
into a disordered 2D conductor is formally equivalent to the DCB in an ultrasmall junction 
with an Ohmic environment. In view of this mapping, we shall use here the standard DCB
theory detailed in Ref.~[\onlinecite{Ingold1992}] to describe the tunneling spectra in 
the normal part of the WL. Within this theory, the tunneling current is given by $I(V) = e 
[\Gamma_{\rm WL \rightarrow tip}(V) - \Gamma_{\rm tip \rightarrow WL}(V) ]$, where the 
inelastic tunneling rates can be expressed as
\begin{eqnarray}
\Gamma_{\rm WL \rightarrow tip}(V) & = & \frac{1}{e^2R_{\rm T}} \int^{\infty}_{-\infty}
dE \int^{\infty}_{-\infty} d\epsilon \, n_{\rm WL}(E) \nonumber \\
& & \times f(E) [1-f(E-\epsilon+eV)] P(\epsilon) \label{eq-rate}
\end{eqnarray}
and $\Gamma_{\rm tip \rightarrow WL}(V) = \Gamma_{\rm WL \rightarrow tip}(-V)$. Here,
$R_{\rm T}$ is the tunneling resistance, $f(E)$ is the Fermi function, $n_{\rm WL}(E)$
is the normalized local DOS of the wetting layer, and $P(E)$ is the probability for an electron
to emit the energy $E$ into the electromagnetic environment \cite{Ingold1992}. Notice
that we have assumed that the tip has an energy-independent DOS. The $P(E)$ function 
is calculated using the total impedance $Z(\omega) = [ i\omega C_{\rm WL} + 1/R_{\rm WL}]^{-1}$, 
where $C_{\rm WL}$ and $R_{\rm WL}$ are the effective capacitance and resistance of the WL, 
respectively, see Fig.~\ref{Fig2}(d). The STM junction capacitance, $C_{\rm T}\lesssim 1$ aF \cite{Brun DCB}, was neglected here.

In Fig.~\ref{Fig2}(c) we show as a solid line the best fit of the experimental ZBA using the 
DCB theory just described. To obtain this fit by means of Eq.~(\ref{eq-rate}),
both $R_{\rm WL}$ and $C_{\rm WL}$ were used as adjustable parameters, while the DOS 
of the WL was assumed to be the one of the non-interacting normal state, \emph{i.e.}\ 
$n_{\rm WL}(E) =1$. The best fit for the experimental temperature was obtained with $C_{\rm WL}=80$ 
aF and $R_{\rm WL}=3.22$ k$\Omega$. The latter value is quite reasonable:  It is close to yet lower 
than the quantum of resistance $h/e^2=25.8$ k$\Omega$, as expected for a correlated 
2D metal manifesting an Altshuler-Aronov ZBA \cite{Altshuler1985}. The $R_{\rm WL}$ value is also consistent with previous \emph{in-situ} transport measurements 
\cite{Pfennigstorf2002}. The value of the capacitance $C_{\rm WL}$ is 
more difficult to interpret. It is about three orders of magnitude higher than the estimated 
self-capacitance $C_{\rm ND} \approx 5\times10^{-2}$ aF of individual Pb nano-domains in Fig.2(b), and almost two orders of magnitude larger than $C_{\rm T}$. This clearly indicates 
that the characteristic length scale involved in the charging effects is much larger than the typical nanometer-size of a domain or of the tunneling junction and thus, is indeed related to the WL. Finally, the quality 
of the fit in Fig.~\ref{Fig2}(c) strongly supports that it is possible to render satisfactorily the tunneling 
data on the basis of the DCB theory using a minimalistic effective RC environment.

\begin{figure}[t]
\includegraphics[width=\columnwidth,clip]{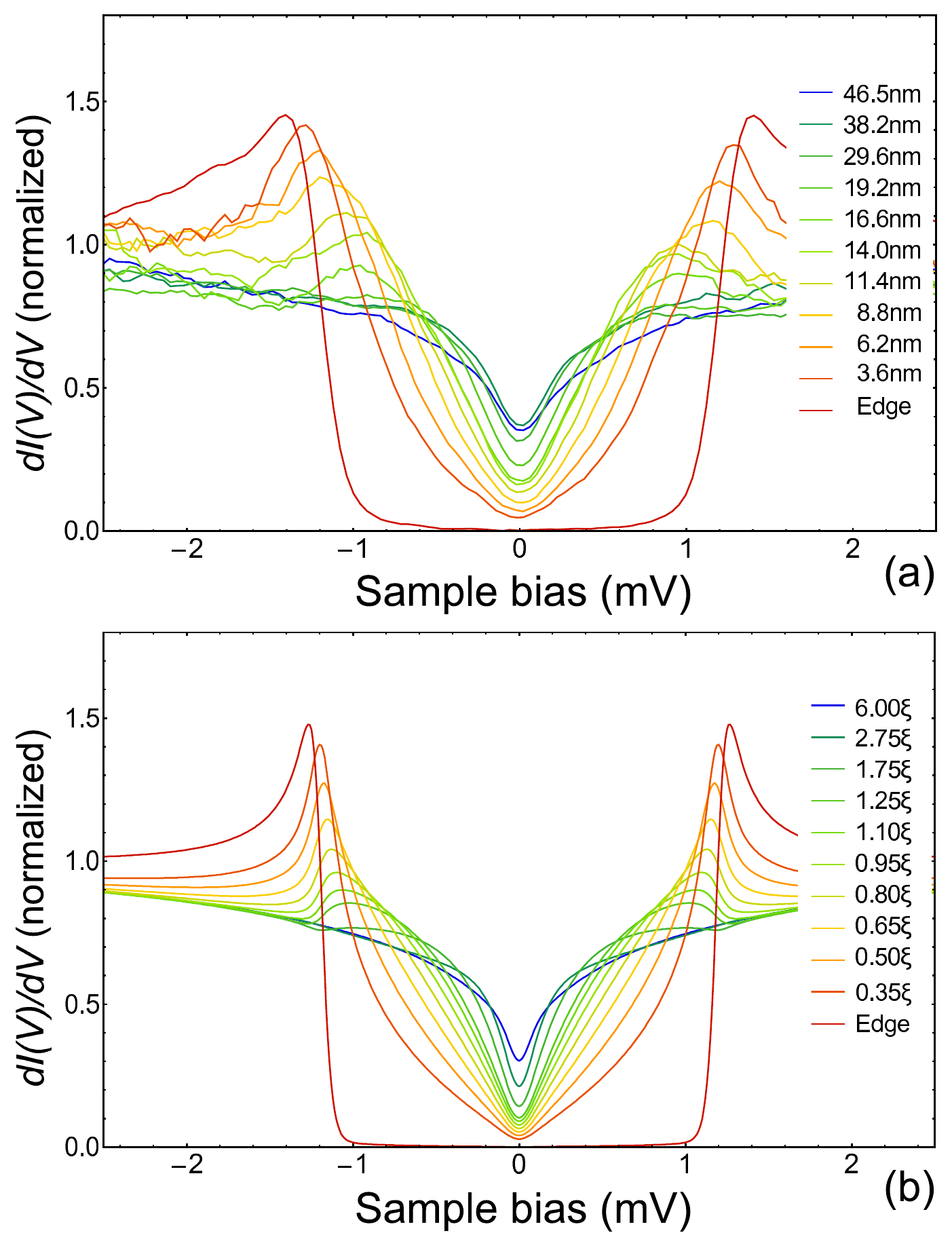}
\caption{(a) Evolution of measured local tunneling conductance spectra (normalized to unity at STS set point $I=0.4$ nA, $V=5$ mV) as a function of the 
distance from the island edge. (b) Computed spectra obtained with the combination of the Usadel 
model and the DCB theory (see text).} \label{Fig3}
\end{figure}

We now turn back to the spatial evolution of the tunneling spectra within the proximity region 
\cite{shadow-effect}. As the tip is moved away from the island and starts probing the WL, the 
tunneling conductance rapidly evolves, see Fig.~\ref{Fig3}(a). At short distances, the peak-to-peak 
separation smoothly reduces, the gap rapidly fills with quasiparticle states at the gap edges, while
at zero-bias $dI(V=0,\textit{\textbf{r}})/dV$ remains low. At larger distances from the island, 
the peak-to-peak separation reduces further, the quasiparticle peaks rapidly smear out and reduce in height.
Moreover, the whole tunneling characteristics inside the gap become more and more V-shaped, and the
conductance at zero bias gradually increases. Finally, 40-50 nm away from the island the local 
$dI(V)/dV$ spectra become peak-less and the ZBA of Fig.~\ref{Fig2}(c) is recovered (for 2D STS maps 
see Ref.~[\onlinecite{SM}]).

The theoretical description of the proximity effect in a correlated normal metal is challenging and, 
although considerable progress has been made in this direction \cite{Oreg1999}, there does
not exist suitable theory that can be directly applied to our hybrid system. In order to explain
our experimental results, and inspired by Ref.~[\onlinecite{Gueron1996}], we shall use here an  
heuristic extension of the DCB theory discussed above. Our assumption is that the proximity
effect can be described by computing the DOS of the WL in Eq.~(\ref{eq-rate}) with the help of
the Usadel equations \cite{Usadel1970}. Let us recall that the Usadel theory has been very 
successful explaining the physics of conventional diffusive SN junctions \cite{Gueron1996,leSueur2008,Belzig1996}.
To be precise, we model our system by means of a SN junction, where N is an infinite normal 
wire and S is an ideal BCS superconducting reservoir with a gap $\Delta$. We also assume 
that the SN interface is perfectly transparent and we neglect the inverse proximity effect 
in the S reservoir (see Ref.~[\onlinecite{SM}] for more details). In this model, the superconducting 
correlations penetrate in the normal metal a distance that depends on energy as $\sqrt{\hbar D/E}$, 
where $D$ is the diffusion constant in the metal. 

To describe the results of Fig.~\ref{Fig3}(a), we use the $P(E)$ function obtained in the analysis
of the normal state of the WL, see Fig.~\ref{Fig2}(e), which is now kept fixed. The superconducting 
gap is also fixed to $\Delta=1.2$ meV from the best fits of the spectra on the Pb island \cite{Nishio,Cren}; the distance to 
the SN interface is measured in units of the length $\xi = \sqrt{\hbar D/\Delta}$. Thus, 
the remaining adjustable parameter is the diffusion constant $D$ entering this expression,
which must have the same value for all spectra. In Fig.~\ref{Fig3}(b) we present a set of numerically 
generated curves; the best match with experimental data is achieved with $\xi=15\pm3$ nm 
resulting in $D \approx 4.1$ cm$^2$/s. Notice that the computed spectra reproduce nicely the above 
discussed spatial evolution of the experimental tunneling data, including even very fine shape 
variations \cite{note}. Moreover, the value of $D$ extracted here is in good agreement with the value 
inferred from \emph{in-situ} conductance measurements of these 1-2 monolayer thick WL \cite{Pfennigstorf2002}. Let us 
emphasize that the diffusion constant in our case is more than one order of magnitude smaller than in 
conventional metals \cite{Gueron1996,leSueur2008}, which confirms that we are dealing with a highly 
disordered system (still away from the metal-insulator transition, $R_{\rm WL}<h/e^2$).

In conclusion, in a very low temperature STM/STS experiment we revealed the proximity effect 
between superconducting Pb islands and an ultimate disordered 2D metal - a non-superconducting 
Pb wetting layer grown on undoped Si(111). We found the tunneling spectra to evolve, on a scale 
of several nanometers, from a superconducting bulk-like behavior inside the islands to a peculiar 
V-shaped zero-bias anomaly far away from them. We attributed this behavior to the interplay between 
disorder and electron correlations in the wetting layer. We developed a phenomenological combination of the DCB theory and the Usadel model to explain these 
observations. Our findings 
not only provide a novel insight into the physics of proximity effect, but they also pave the way 
for studying new aspects of this effect in a great variety of hybrid systems including \emph{in-situ} 
fabricated lateral SNS junctions in which the correlations play an important role.

We thank Hermann Grabert for useful discussions. This work was supported by grants from the 
University Pierre et Marie Curie (UPMC) ``Emergence" and by CNRS PhD grant (L.S.-G.).
J.C.C. and F.S.B. acknowledge financial support from the Spanish MICINN (Contracts No. 
FIS2011-28851-C02-01 and FIS2011-28851-C02-02).


\begin{thebibliography}{00}
\bibitem{Non-supercond WL}
By applying an external magnetic field up to 6 T we additionally checked the absence 
of vortices in the WL and, more importantly, a general insensitivity of the ZBA spectra 
to the magnetic field.
\bibitem{shadow-effect}
Note that due to the finite size of the STM tip apex, the WL in the vicinity of the 
Pb-island edge is not accessible by STS. Within this ``shadow" region of a few nm the 
tip does not probe the WL but rather the upper edge of the island (see Ref.~[\onlinecite{SM}]).
\bibitem{note}
Notice, however, that the agreement is not perfect, which shows the limitations of 
our simple model where a fixed RC environment was used to describe all the experimental 
curves of Fig.~\ref{Fig3}(a).
\end{thebibliography}
\end{document}